\documentclass[10pt,romanappendices,conference,twocolumn,letter]{IEEEtran}
\IEEEoverridecommandlockouts
\usepackage{cite}
\usepackage{graphicx, amsfonts,amsmath, enumerate, amsthm, amssymb,etaremune,array,url}
\usepackage{subfigure,setspace,multirow,psfrag}
\graphicspath{{../figures/}}
\usepackage{subfigure}

\newcommand{\Lambdav}{\mbox{\boldmath$\Lambda$}}

\newcommand{\nuv}{\mbox{\boldmath$\nu$}}
\newcommand{\omegav}{\mbox{\boldmath$\omega$}}

\newtheorem{corollary}{Corollary}

\newtheorem{proposition}{Proposition}
\newtheorem{lemma}{Lemma}
\newtheorem{assumption}{Assumption}

\begin{document}

\title{Massive MIMO with a Generalized Channel Model: Fundamental Aspects
\thanks{The work of M. Matthaiou and H. Tataria was supported by EPSRC,
UK, under grant EP/P000673/1. The work of P. J. Smith was supported
by the Royal Academy of Engineering, UK, via the Distinguished Visiting
Fellowship DVF1617/6/29.}}

\author{\IEEEauthorblockN{Michail Matthaiou\IEEEauthorrefmark{1}, Hien Quoc Ngo\IEEEauthorrefmark{1}, Peter J. Smith\IEEEauthorrefmark{2}, Harsh Tataria\IEEEauthorrefmark{1}, and Shi Jin\IEEEauthorrefmark{3}}
\IEEEauthorblockA{$^*$Institute of Electronics, Communications and Information Technology (ECIT), Queen's University Belfast, Belfast, U.K.}
\IEEEauthorblockA{$^\dagger$School of Mathematics and Statistics, Victoria University of Wellington, Wellington, New Zealand}
\IEEEauthorblockA{$^\ddagger$National Mobile Communications Research Laboratory, Southeast University, Nanjing, P. R. China\\}
Email: \{m.matthaiou, hien.ngo, h.tataria\}@qub.ac.uk, peter.smith@vuw.ac.nz, jinshi@seu.edu.cn}
\maketitle
\begin{abstract} Massive multiple-input multiple-output (MIMO) is becoming a mature technology, and has been approved for standardization in 
the 5G ecosystem. Although there is a large body of papers on the theoretical analysis of massive MIMO, the majority of relevant work assumes the simplified, yet overly idealistic, Kronecker-type model for spatial correlation. Motivated by the deficiencies of the Kronecker model, we invoke a naturally generalized spatial correlation model, that is the Weichselberger model.
For this model, we pursue a comprehensive analysis of massive MIMO performance in terms of channel hardening and favorable propagation (FP).
We identify a number of scenarios under which massive MIMO may fail, in terms of channel hardening and FP, and discuss their relevance from a practical perspective. 
\end{abstract}

\section{Introduction}
Massive MIMO is not simply a promising technology but, as a matter of fact, it is making 
its way into 5G trials and standards \cite{MAMI_book,Ngo1}. The biggest body of related literature
is based on Kronecker-type models, which offer analytical tractability, thereby, facilitating
performance analysis and transceiver design \cite{Kronecker}. Yet, it is a well-known fact from the early days of conventional MIMO,
that Kronecker-based models enforce the spatial correlation properties at both ends to be separable \cite{Kronecker_def};
that is all DoAs are treated as completely independent from the DoDs, and vice versa
This implication produces artifact paths lying at the vertical and horizontal intersections of the real DoD and DoA spectral
peaks \cite{Kronecker2}. These artifacts increase the apparent diversity but decrease the apparent capacity since they take away energy from all real paths that do not lie at the intersection points so that the overall power is kept constant. 

A far more realistic channel model is the so-called Weichselberger model \cite{Weich_model}, 
which alleviates the deficiencies of the Kronecker model by considering the joint correlation structure of both ends; 
therefore, the average coupling between the spatial subchannels is effectively modeled. We also note that the
Weichselberger model includes the Kronecker model and the virtual channel representation (VCR) \cite{Sayeed} as special cases. 
Despite its importance, there is a dearth of literature on the performance analysis of conventional multi-user MIMO (MU-MIMO) and massive MIMO with this generalized Gaussian fading model. This can be partially attributed to the increased number of parameters that have to be specified
for the Weichselberger model compared to the Kronecker model and the VCR. In the MU-MIMO literature, we point out the work 
of \cite{XGao}, which investigated the capacity-achieving input covariance matrix for
a single-user Weichselberger Ricean fading MIMO channel and \cite{Tulino} which extended \cite{XGao} to the MU case but only in the low-power regime. More recently, \cite{CKWen} investigated the asymptotic sum-rate of the MU Weichselberger Ricean fading
MIMO channel using the replica method. 

The massive MIMO literature is even more scarce with respect to the Weichselberger model. To the best of our knowledge, the only relevant works are \cite{LTH_thesis, Aissa}. The work in \cite{LTH_thesis} compared the accuracy of the Kronecker and Weichselberger models against a set of massive MIMO measurement data at 2.6 GHz and concluded that the latter indeed provides  more accurate modeling.
On the other hand, \cite{Aissa} considered both centralized and distributed massive MIMO networks and evaluated their performance 
in terms of spectral efficiency. Their theoretical analysis was entirely based on the Weichselberger model but neglected any line-of-sight components (i.e., Rayleigh fading conditions were assumed). Another family of papers was developed by Raghavan \textit{ et al.} (see \cite{Raghavan} and references therein) which formulated a mathematical framework to encompass the different types of correlation-based models and examined how the level of sparsity affects their capacity performance using tools of random matrix theory.

This paper moves away from the state-of-art and analyzes, for the first-time, the theoretical performance of massive MIMO using the 
Weichselberger model. After introducing the new system model and discussing some basic statistical properties, our analysis targets the two fundamental performance metrics of any massive MIMO communication system, namely \textit{channel hardening} and \textit{favorable propagation}.
By leveraging tools of Gaussian theory, we derive mathematical conditions under which these two concepts become valid. We also identify analytically scenarios under which these two concepts break down and corroborate them with a set of numerical results. Our work complements and extends some recent theoretical papers on massive MIMO with pure LoS fading \cite{Ngo_EUSIPCO,Masouros}, i.i.d Rayleigh fading \cite{Ngo_EUSIPCO}, semi-correlated Rayleigh fading \cite{GLOBECOM}, and semi-correlated Ricean fading \cite{Tataria2017,Matthaiou_WCL}. 

\textit{Notation:} We use upper and lower case boldface to denote matrices and vectors, respectively. The $n\times n$ identity matrix is expressed as $\mathbf{I}_{n}$. A complex normal vector with mean ${\mathbf b}$ and covariance ${ \bf \Sigma}$ reads as $\mathcal{CN}({ \mathbf b},{ \bf \Sigma})$. The expectation of a random variable is denoted as ${\mathbb E}\left[\cdot \right] $, while the matrix trace by $\mathrm{{tr}(\cdot )}$. The symbols $\left( \cdot \right) ^{*}$, $\left( \cdot \right) ^{T}$ and $\left(\cdot \right) ^{H}$ represent the conjugate, transpose and Hermitian transpose of a matrix. The notation $\stackrel{{\rm a.s.}}{\longrightarrow}$ implies almost sure convergence, $\stackrel{{\rm P}}{\longrightarrow}$ denotes convergence in probability, while $\odot$ denotes the element-wise (Hadamard) multiplication between two matrices (or vectors). We also introduce the notation $\underline{{\bf X}}\triangleq{\bf X}\odot{\bf X}^*$ and also invoke $|| {\bf x}||_\infty=\max\left(|x_1|,|x_2|,\ldots\right)$ and $||{\bf X}||_{\textrm{max}}=\max\limits_{ij}|x_{ij}|$.

\section{System model}
We consider the uplink of a massive MIMO system, where the BS is equipped with $M$ antennas and serves $L$ single-antenna
users, where $M\gg L$. The $M\times 1$ channel from the $k$-th user to the BS is 
\cite{Weich_model}
\begin{align}
	{\bf h}_k & = \eta_k\bar{\bf {h}}_k+\gamma_k\underbrace{{\bf U}_k\left(\widetilde{{\omegav}}_k\odot{\bf h}_{{\tt iid}}\right)}_{\triangleq \widetilde{\bf {h}}_k}
\label{eq:System_model1} 
\end{align}
where $\bar{\bf {h}}_k$ is the LoS component, $\eta_k\triangleq (K_k/(K_k+1))^{1/2}$ and $\gamma_k\triangleq(1/(K_k+1))^{1/2}$, with $K_k$ being the Ricean $K$-factor seen by the $k$-th user. Moreover, $\widetilde{{\omegav}}_k$ is the element-wise square root of ${{\omegav}}_k=[\omega_{k,1},\ldots\omega_{k,M}]^T\in \mathbb{R}^{M\times 1}$ (which will be defined shortly), while $\bf {h}_{{\tt iid}}$ is an $M\times 1$ vector with i.i.d. $CN(0,1)$ entries. We also define the small-scale fading matrix ${\bf H}\triangleq[{\bf h}_1,{\bf h}_2,\ldots,{\bf h}_L]\in {\mathbb C}^{M\times L}$. The spatial structure of the random component in \eqref{eq:System_model1} can be efficiently captured by the one-sided correlation matrix 
\begin{align}
	{\bf Q}_k &= {\mathbb E}\left[ \widetilde{\bf {h}}_k\widetilde{\bf {h}}_k^H\right]= {\bf U}_k{\mathbb E}\left[\left(\widetilde{{\omegav}}_k\odot{\bf h}_{{\tt iid}}\right)\left(\widetilde{{\omegav}}_k\odot{\bf h}_{{\tt iid}}\right)^H  \right]{\bf U}_k^H\notag \\
	&= {\bf U}_k\left(\widetilde{{\omegav}}_k\widetilde{{\omegav}}_k^T\right)\odot{\mathbb E}\left[{\bf h}_{{\tt iid}}{\bf h}_{{\tt iid}}^H\right]{\bf U}_k^H = {\bf U}_k \Lambdav_k {\bf U}_k^H
	\label{eq:Covariance}
\end{align}
where $\Lambdav_k = \textrm{diag}\left({\omegav}_k\right)$ and we have leveraged the following property $({\bf a}\odot{\bf b}) ({\bf c}\odot{\bf d})^H = \left({\bf a}{\bf c}^H\right)\odot\left({\bf b}{\bf d}^H\right)$. Looking at \eqref{eq:Covariance}, we infer that 
${\bf U}_k$ contains the eigenbases at the BS side related with user $k$. The diagonal matrix $\Lambdav_k$ contains the eigenvalues of 
${\bf Q}_k$ which are also equal to the real coefficients $\omega_{k,1},\ldots\omega_{k,M}$. As was articulated in 
\cite{Weich_model}, $\omega_{m,\ell}$ represents the coupling coefficients which specify the mean amount of energy coupled from the $\ell$-th user to the $m$-th receive eigenvector. Referring back to \eqref{eq:System_model1}, the following constraints should be satisfied: $\bar{\bf {h}}_k^H\bar{\bf {h}}_k=M$ and ${\rm tr}(\Lambdav_k)=M, \forall k=1,\ldots,L$ \cite{CKWen, Tataria2017}. 


\section{Channel hardening and FP}
Since its invention by Marzetta \cite{Marzetta}, the theoretical advancement of massive MIMO has been based on two fundamental concepts: 
\textit{channel hardening} and \textit{FP}. In this section, we will elaborate on these concepts and identify scenarios under which they break down. Such scenarios are particularly important as they correspond to environments which either increase the randomness of an uplink channel or boost the inter-user interference. To keep our notation clean, we focus on the perfect CSI case, though very similar results can be obtained for the imperfect CSI case. The following results will be rather useful in the subsequent analysis:
\begin{proposition}
For the channel model in \eqref{eq:System_model1}, we have that 
\begin{align}
	{\mathbb E}\Big[||{\bf h}_k||^4\Big] &= \gamma_k^4{\rm tr}\left(\Lambdav_k^2\right)
	+M^2+2\eta_k^2\gamma_k^2\Big({\bf v}_k^H {\Lambdav}_k {\bf v}_k \Big)	\label{eq:Signal_power1}\\
	{\mathbb E}\Big[||{\bf h}_k||^2\Big] & = M,\label{eq:Signal_power2}
	\end{align}
	where ${\bf v}_k = {\bf U}_k^H\bar{\bf {h}}_k=[\nu_{k,1},\ldots,\nu_{k,M}]^T$.
\begin{proof}
The proof is provided in Appendix \ref{app1}.
\end{proof}
\label{Moments}
\end{proposition}

\begin{proposition}\label{Moments2}
For the channel model in \eqref{eq:System_model1}, we have that 
\begin{align}
	{\mathbb E}\Big[\big|{\bf h}_k^H{\bf h}_\ell\big|^2\Big] &= \eta_k^2\eta_\ell^2\big|\bar{\bf {h}}_k^H\bar{\bf {h}}_\ell\big|^2
	+\gamma_k^2\gamma_\ell^2{\rm tr}\Big({\bf Q}_k {\bf Q}_\ell	\Big)\notag\\
	&+\eta_k^2\gamma_\ell^2\Big(\bar{\bf {h}}_k^H {\bf Q}_\ell\bar{\bf {h}}_k \Big)
	 + \gamma_k^2\eta_\ell^2\Big(\bar{\bf {h}}_\ell^H {\bf Q}_k\bar{\bf {h}}_\ell \Big).
	\label{eq:Int_power}
\end{align}
\begin{proof}
The proof is provided in Appendix \ref{app2}.
\end{proof}
\end{proposition}
It is noteworthy that the results in Propositions 1 and 2 can generalize our results in \cite{Tataria2017} for the case of semi-correlated Ricean fading with Kronecker-type of spatial correlation. The following result will be particularly useful in our asymptotic analysis:
\begin{lemma}
Tchebyshev's theorem: Let $X_1, X_2,\ldots, X_n$  be independent RVs with ${\mathbb E}\left[X_i \right] = \mu_i$ and
$\mathrm {Var}\left(X_i\right)\leq z< \infty, \forall i\in{1,2,\ldots,n}.$ Then, as $n\rightarrow\infty$
\begin{align*}
\frac{1}{n}\left(X_1+X_2+\ldots+X_n\right)-\frac{1}{n}\left(\mu_1+\mu_2+\ldots+\mu_n\right)\stackrel{{\rm P}}{\longrightarrow}0.
\end{align*}
\end{lemma}

\subsection{Channel hardening}
Channel hardening arises whenever the randomness of a fading channel is averaged, converting the normalized channel power into a deterministic quantity. The authors of \cite{Emil_book} provided the following definition of asymptotic channel hardening
\begin{align}
	\frac{||{\bf h}_k||^2}{{\mathbb E}\left[||{\bf h}_k||^2\right]}\stackrel{{\rm a.s.}}{\longrightarrow}  1,~{\rm as}~M\rightarrow\infty.
\end{align}

\begin{assumption}
As $M\rightarrow\infty$, for every $k=1,\ldots,L$, $\limsup\limits_{M}||\underline{\omegav_k} ||_\infty< \infty$ and $\limsup\limits_{M}||\underline{{\bf v}_k}||_\infty< \infty$. 
\end{assumption}

The physical interpretation of Assumption 1 is that no $\omega_{k,m}, |\nu_{k,m}|, m=1,\ldots,M$ grows without bound as $M$ increases.
We will later on see what happens when this constraint is not satisfied. 
\begin{corollary}
For the channel model in \eqref{eq:System_model1} and provided that Assumption 1 is fulfilled, we have 
\begin{align}
		\frac{||{\bf h}_k||^2}{{\mathbb E}\left[||{\bf h}_k||^2\right]}\stackrel{{\rm P}}{\longrightarrow}  1,~{\rm as}~M\rightarrow\infty.
\label{eq:CH}
\end{align}
\begin{proof}
The proof begins by evaluating the asymptotic behavior of the term $\widetilde{\bf {h}}_k^H\widetilde{\bf {h}}_k$ which arises in the expansion of $||{\bf h}_k||^2$. In particular, we get
\begin{align}
	\frac{1}{M}\widetilde{\bf {h}}_k^H\widetilde{\bf {h}}_k &= \frac{1}{M}{\rm tr}\left(\left(\widetilde{{\omegav}}_k\widetilde{{\omegav}}_k^T\right)\odot\left({\bf h}_{{\tt iid}}{\bf h}_{{\tt iid}}^H\right)\right)\notag\\
	&=\frac{1}{M}\sum_{m=1}^M \omega_{k,m}|X_m|^2\stackrel{{\rm P}}{\longrightarrow}\frac{1}{M}\sum_{m=1}^M \omega_{k,m}=1
	\label{eq:CH2}
\end{align}
where $X_m$ is an i.i.d. complex Gaussian RV with zero mean and unit variance. Note that in the last step of \eqref{eq:CH2}, we have utilized Lemma 1. In a very similar manner, we can show that $\frac{1}{M}\widetilde{\bf {h}}_k^H\bar{\bf {h}}_k\stackrel{{\rm P}}{\longrightarrow}0$ and $\frac{1}{M}\bar{\bf {h}}_k^H\widetilde{\bf {h}}_k\stackrel{{\rm P}}{\longrightarrow}0$. The proof then concludes after some basic algebra. 
\end{proof}
\end{corollary}

\textit{Discussion:} By inspection of \eqref{eq:CH2}, we can infer that channel hardening will break down if some $\omega_{k,m}, 1\leq n \leq M$ scale as ${\mathcal O}(M^\epsilon)$ with $0 <\epsilon\leq 1$. In this case, $\frac{1}{M}\widetilde{\bf {h}}_k^H\widetilde{\bf {h}}_k$ will asymptotically start to behave as a linear combination of $|X_m|^2$ terms, which are exponential variates, thereby experiencing random fluctuations. From a practical perspective, this scenario kicks in if the energy from the $k$-th user couples only into a small number of receive eigenvectors. This can happen when there is a limited number of resolvable multipath components from the $k$-th user impinging on the BS. Similar conclusions can be drawn for the elements of ${\bf v}_k$ that appear in the expansion of $\frac{1}{M}\widetilde{\bf {h}}_k^H\bar{\bf {h}}_k$ and
$\frac{1}{M}\bar{\bf {h}}_k^H\widetilde{\bf {h}}_k$.

Mathematically speaking, the convergence in probability in Corollary 1 is a weaker condition compared to almost sure convergence. Nevertheless, in a practical system such a difference will cause little performance variation (if any). This observation is corroborated by our numerical results. Most important though is the deviation from these ``asymptotically-optimal'' conditions in the finite number of antennas regime; this can be quantified by the scaled second-order moment of the channel gain, defined as below
\begin{align}
\mathrm {Var}\left(\frac{||{\bf h}_k||^2}{{\mathbb E}\left[||{\bf h}_k||^2\right]}\right)=\frac{\gamma_k^2}{M^2}
\Big(\underbrace{2\eta_k^2{\bf {v}}_k^H \Lambdav_k {\bf {v}}_k}_{\textrm{LoS contribution}} +\underbrace{\gamma_k^2{\rm tr}\left(\Lambdav_k^2\right)\Big)}_{\textrm{NLoS contribution}}.
\label{eq:Variance}
\end{align}
The above expression decouples nicely the NLoS and LoS contributions; in fact, the scaled variance of the signal power is an increasing function of $\gamma_k$; for $\gamma_k=0$ (pure LoS propagation), this variance becomes exactly zero regardless of the value of $M$, whereas for $\gamma_k=1$ (correlated Rayleigh fading conditions), the right-hand side of \eqref{eq:Variance} 
becomes equal to ${\rm tr}\left(\Lambdav_k^2\right)/M^2$. This value agrees with \cite[Eq. (2.17)]{Emil_book}. For an arbitrary and fixed value of $\gamma_k\in[0,1]$, we can make the following observations:
\begin{itemize}
	\item From the Rayleigh-Ritz theorem, the LoS component in \eqref{eq:Variance} is upper and lower bounded as follows:
	\begin{align}
	\frac{\min({\omegav}_k)}{M}\leq\frac{1}{M^2}{\bf {v}}_k^H{\Lambdav}_k	{\bf {v}}_k \leq \frac{\max({\omegav}_k)}{M}\leq 1.
	\label{eq:Ritz}
\end{align}
Interestingly, the lower limit above does not go asymptotically to zero when $\bar{\bf {h}}_k^H$ is aligned with the weakest eigenvector of ${\bf Q}_k$ and the corresponding eigenvalue scales as ${\cal O}(M)$. More generally, 
by definition $\bar{\bf {h}}_k $ can be expressed as a linear combination of the linearly independent eigenvectors
of ${\bf Q}_k$ (i.e. the columns of ${\bf U}_k$). If the corresponding eigenvalues scale as ${\cal O}(M)$, then $\left({\bf {v}}_k^H{\Lambdav}_k	{\bf {v}}_k\right)/M^2$ does not vanish as $M\rightarrow\infty$. As such, whenever we have non-vanishing alignment of a LoS response of the $k$-th user with the dominant eigenvectors of the corresponding covariance matrix ${\bf Q}_k$, asymptotic channel hardening becomes challenging. 
	\item The NLoS component in \eqref{eq:Variance} is maximized when ${\bf Q}_k$ is rank-1\cite{Emil_book}, in which case ${\rm tr}\left(\Lambdav_k^2\right)=M^2$. Then,
	 \begin{align*}
\mathrm {Var}\left(\frac{||{\bf h}_k||^2}{{\mathbb E}\left[||{\bf h}_k||^2\right]}\right)=\frac{\gamma_k^2}{M^2}\left(2\eta_k^2M|\bar{\bf {h}}_k^H {\bf u}_{k,1} |^2+\gamma_k^2M^2\right)
\end{align*}
where ${\bf u}_{k,1}$ is the dominant eigenvector of ${\bf Q}_k$. On the other hand, the variance is minimized when ${\bf Q}_k={\bf I}_M$ in which case ${\rm tr}\left(\Lambdav_k^2\right)=M$ 
and \eqref{eq:Variance} becomes
 \begin{align*}
\mathrm {Var}\left(\frac{||{\bf h}_k||^2}{{\mathbb E}\left[||{\bf h}_k||^2\right]}\right)=\frac{\gamma_k^2}{M}\left(1+\eta_k^2\right),
\end{align*}
which goes smoothly to zero as $M\rightarrow\infty$.
\end{itemize}

\begin{figure}
\centering\includegraphics[width=8.85cm]{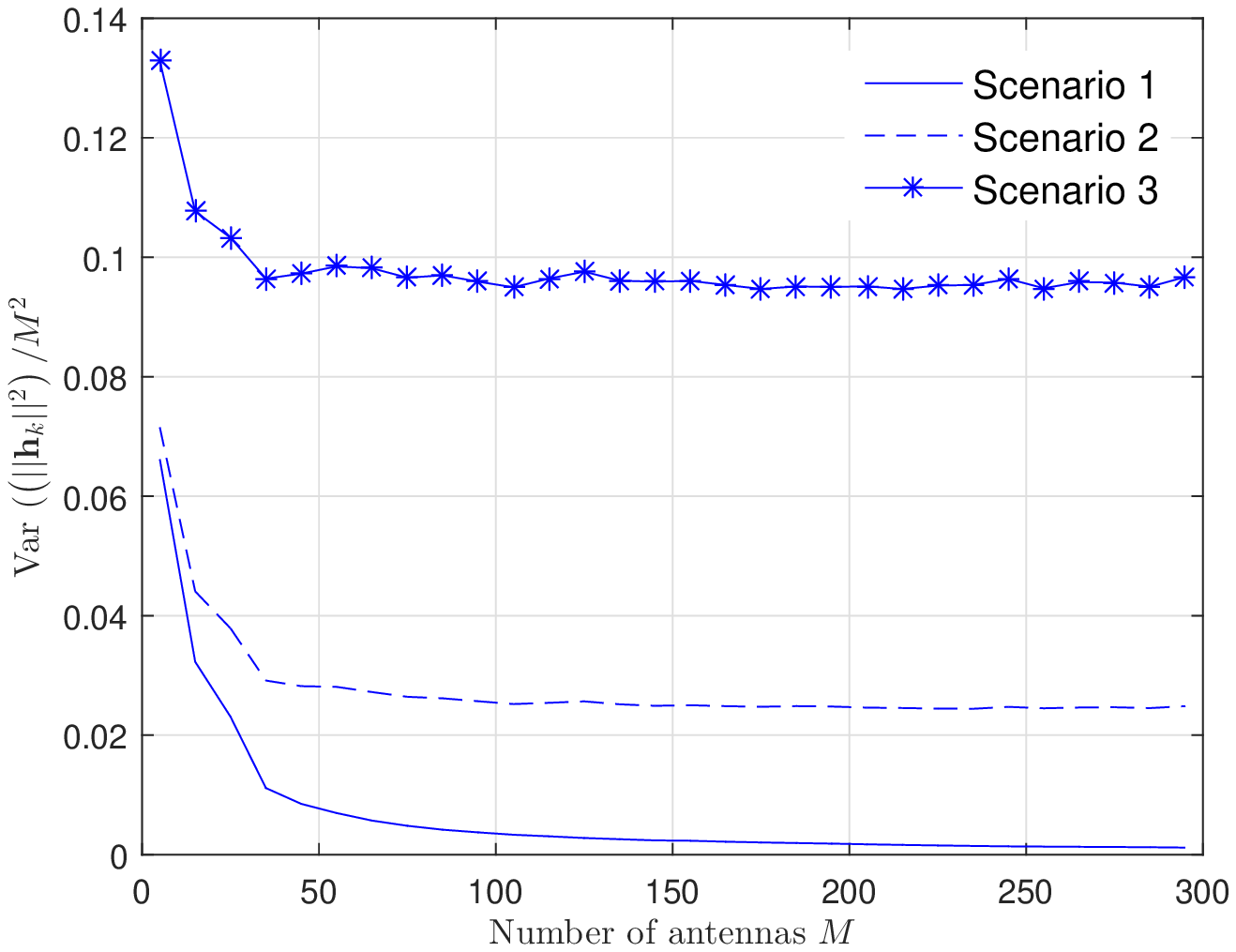}
\caption{$\mathrm {Var}\left({||{\bf h}_k||^2}\right)/M^2 $ against the number of antenna, $M$ ($K_k=0.5, \phi_k=\pi/3$).}
\label{fig:Hardening}
\end{figure}

We will now investigate how the variance in \eqref{eq:Variance} behaves for a range of scenarios. Assuming a uniform linear array at the BS, the LoS channel response can be expressed as
\begin{align}
	\bar{\bf {h}}_k = \left[1, e^{j2\pi d\cos(\phi_k)},\ldots,e^{j2\pi d(M-1)\cos(\phi_k)}\right]^T
	\label{eq:LoS_resp}
\end{align} 
where $d$  is the equidistant inter-element antenna spacing normalized by the carrier wavelength
and $\phi_k$ is the angle-of-arrival from the $k$-user. We will now consider three different scenarios for the
coupling vector ${\omegav}_k$ of the $k$-th user:
\begin{align*}
	&\textrm{Scenario 1:}~ {\omegav}_k = [1,1,\ldots,1]^T\\
	&\textrm{Scenario 2:}~ {\omegav}_k = [M/2,M/(2M-2),\ldots,M/(2M-2)]^T\\
	&\textrm{Scenario 3:}~ {\omegav}_k = [M,0,\ldots,0]^T.
\end{align*}
The first scenario represents an equal distribution of power across all antenna elements and resembles that of i.i.d. Rayleigh fading; in the second and third scenarios, there are entries $\omega_{k,m}$ that scale as ${\cal O}(M)$. In particular, the last scenario represents
an extreme case where only a single antenna can capture energy from the $k$-th user. In Fig. \ref{fig:Hardening},
we plot $\mathrm {Var}\left({||{\bf h}_k||^2}/{{\mathbb E}\left[||{\bf h}_k||^2\right]}\right)=\mathrm {Var}\left({||{\bf h}_k||^2}\right)/M^2$ against $M$. Note that the unitary matrices ${\bf U}_k$ have been randomly generated. 
The figure validates that for isotropic fading conditions (as those studied in the early papers on massive MIMO \cite{Ngo1,Ngo_EUSIPCO}), the normalized variance converges smoothly to zero. This is because all entries $\omega_{k,m}$ scale as ${\cal O}(1)$. On the other hand, for Scenarios 2 and 3, the findings are substantially different.
As a matter of fact, it is observed that the scaled variance converges to non-zero limits, thereby indicating the deviation from the channel hardening regime. The situation is exacerbated for Scenario 3, which corroborates our theoretical analysis that rank-1 matrices maximize the NLoS contribution in the variance expression in \eqref{eq:Variance}.

\subsection{Favorable propagation}
We now turn our attention to FP, which has been identified as the 
key feature of massive MIMO that enables successful inter-user interference cancellation \cite{Ngo_EUSIPCO,Masouros,GLOBECOM}. Mathematically speaking, two channel vectors offer asymptotic FP if they satisfy the following relationship \cite{Emil_book}
\begin{align}
	\frac{{\bf h}_k^H{\bf h}_\ell}{\sqrt{{\mathbb E}\left[||{\bf h}_k||^2\right]{\mathbb E}\left[||{\bf h}_\ell||^2\right]}}\stackrel{{\rm a.s.}}{\longrightarrow}  0,~{\rm as}~M\rightarrow\infty.
	\label{eq:FP}
\end{align}

Let us now define the following matrices: ${\bf\Omega}_{\ell,k}=\omegav_\ell\omegav_k^T$
and ${\bf R}_{k,\ell}={\bf U}_k^H{\bf U}_\ell$ and the vector $\nuv_{k,\ell}=\bar{\bf {h}}_k^H{\bf U}_\ell$
\begin{assumption}
As $M\rightarrow\infty$, for every $k,\ell=1,\ldots,L$, $\limsup\limits_{M}||\underline{{\bf\Omega}_{\ell,k}}||_{\textrm{max}}< \infty$ and $\limsup\limits_{M}||\underline{{\bf R}_{k,\ell}}||_{\textrm{max}}< \infty$. 
\end{assumption}

\begin{assumption}
As $M\rightarrow\infty$, for every $k,\ell=1,\ldots,L$, $\limsup\limits_{M}||\underline{\nuv_{k,\ell}}||_\infty< \infty$.
\end{assumption}

\begin{assumption}
As $M\rightarrow\infty$, for every $k,\ell=1,\ldots,L$, $\limsup\limits_{M}|\bar{\bf {h}}_k^H\bar{\bf {h}}_\ell|< \infty$.
\end{assumption}

As with Assumption 1, Assumption 2-4 bound every individual entry of the involved vectors/matrices to remain finite, as the number of antennas, $M$, grow large. We will later on see what happens if any of these Assumptions are violated. 
\begin{corollary}
For the channel model in \eqref{eq:System_model1} and provided that Assumptions 1-4 are fulfilled, we have
\begin{align}
\frac{{\bf h}_k^H{\bf h}_\ell}{\sqrt{{\mathbb E}\left[||{\bf h}_k||^2\right]{\mathbb E}\left[||{\bf h}_\ell||^2\right]}}\stackrel{{\rm P}}{\longrightarrow}  0,~{\rm as}~M\rightarrow\infty.
\label{eq:FP1}
\end{align}
\begin{proof}
By expanding the left-hand side of \eqref{eq:FP} we obtain 
\begin{align*}
	&\frac{{\bf h}_k^H{\bf h}_\ell}{\sqrt{{\mathbb E}\left[||{\bf h}_k||^2\right]{\mathbb E}\left[||{\bf h}_\ell||^2\right]}}=\frac{1}{M}\big(\eta_k \bar{\bf {h}}_k^H+\gamma_k \widetilde{\bf {h}}_k^H\big)\big(\eta_\ell \bar{\bf {h}}_\ell+\gamma_\ell \widetilde{\bf {h}}_\ell\big)\notag\\
&=	\frac{1}{M}\big(\eta_k\eta_\ell \bar{\bf {h}}_k^H\bar{\bf {h}}_\ell + \eta_k\gamma_\ell \bar{\bf {h}}_k^H\widetilde{\bf {h}}_\ell+\gamma_k\eta_\ell \widetilde{\bf {h}}_k^H\bar{\bf {h}}_\ell+\gamma_k\gamma_\ell\widetilde{\bf {h}}_k^H\widetilde{\bf {h}}_\ell\big).
\end{align*}
Now we can apply Lemma 1 on each random term of the above equation, which requires Assumptions 1-3 to be fulfilled. The deterministic quantity $\frac{1}{M}\bar{\bf {h}}_k^H\bar{\bf {h}}_\ell$ remains bounded if and only if Assumption 4 is satisfied. This concludes the proof.
\end{proof}
\end{corollary}

An important measure of performance is how orthogonal the channel vectors are for a practical number of antennas; this, in turn, will also indicate the actual amount of inter-user interference. This can be quantified by the variance of \eqref{eq:FP}, that is
\begin{align}
\mathrm {Var}\left(\frac{{\bf h}_k^H{\bf h}_\ell}{\sqrt{{\mathbb E}\left[||{\bf h}_k||^2\right]{\mathbb E}\left[||{\bf h}_\ell||^2\right]}}\right)=\frac{1}{M^2}{\mathbb E}\Big[\big|{\bf h}_k^H{\bf h}_\ell\big|^2\Big].
\label{eq:Var_FP}	
\end{align}
Recall that the right-hand side of \eqref{eq:Var_FP} has already been derived in closed-form in \eqref{eq:Int_power}.
In some of our recent work \cite{Matthaiou_WCL}, we identified scenarios under which terms that have the form of \eqref{eq:Int_power} do not asymptotically go to zero. In a nutshell, this happens whenever we have strong alignment of
two distinct LoS responses and/or non-vanishing alignment of a LoS response of the $k$-th user with the eigenvectors of
the covariance matrix of the $\ell$-th user, i.e. ${\bf Q}_\ell$, whose eigenvalues scale as ${\cal O}(M)$. 

Let us see this via a toy example and focus on the term $\bar{\bf {h}}_\ell^H {\bf Q}_k\bar{\bf {h}}_\ell$. 
The unitary eigenvector matrix of ${\bf Q}_k$ can be expressed through its column eigenvectors as follows ${\bf U}_k=\left[{\bf u}_1^{(k)},{\bf u}_2^{(k)},\ldots,{\bf u}_M^{(k)}\right]$. Now, we consider the case where $\bar{\bf {h}}_\ell$ is a linear combination of the two principal eigenvectors of ${\bf U}_k$, such that\footnote{Note that $\bar{\bf {h}}_\ell$ can be \textit{de facto} expressed as a linear combination of ${\bf u}_m^{(k)}, m=1,\ldots,M$, since they fill the complex space ${\mathbb C}^M$.}
\begin{align*}
	\bar{\bf {h}}_\ell = \sqrt{\frac{M}{2}}{\bf u}_1^{(k)}+\sqrt{\frac{M}{2}}{\bf u}_2^{(k)}.
\end{align*}
We can easily show after some basic algebra that
\begin{align}
	\frac{1}{M^2}\bar{\bf {h}}_\ell^H {\bf Q}_k\bar{\bf {h}}_\ell=\frac{\omega_{k,1}+\omega_{k,2}}{2M}.
	\label{eq:quad}
\end{align}
Thus, if any of the eigenvalues $\omega_{k,1}, \omega_{k,2}$ scales as ${\cal O}(M)$, FP will break down. 

\begin{figure}
\centering\includegraphics[width=8.85cm]{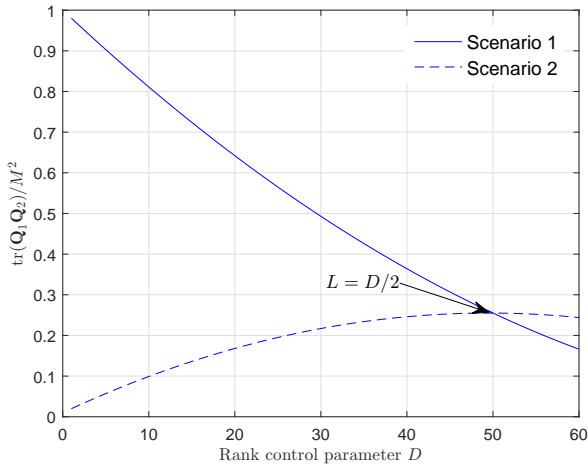}
\caption{Inter-user covariance interference term, ${\rm tr}({\bf Q}_1{\bf Q}_2)/M^2$ against the  rank control parameter ($M=100$).}
\label{fig:traces}
\end{figure}

We can now investigate further the behavior of  the \textit{inter-user covariance  interference} term ${\rm tr}\Big({\bf Q}_k {\bf Q}_\ell\Big)$ in \eqref{eq:Int_power}. Our analysis begins by defining ${\bf V}_{k\ell} = {\bf U}_\ell^H{\bf U}_k$ and then expanding the trace term as follows:
\begin{align}
	{\rm tr}\Big({\bf Q}_k {\bf Q}_\ell\Big) &= {\rm tr}\Big({\bf U}_k\Lambdav_k{\bf U}_k^H {\bf U}_\ell\Lambdav_\ell{\bf U}_\ell^H\Big)\notag\\
	&={\rm tr}\Big(\Lambdav_\ell{\bf V}_{k\ell}\Lambdav_k{\bf V}_{k\ell}^H\Big)\notag\\
&=\sum_{m=1}^M\sum_{n=1}^M\omega_{\ell,m}\omega_{k,n}|{\bf V}_{k\ell}(m,n)|^2\notag\\
&\leq ||{\bf V}_{k\ell}||^2_{\textrm{max}} \sum_{m=1}^M\sum_{n=1}^M \omega_{\ell,m}\omega_{k,n}\notag\\
&\leq M^2 ||{\bf V}_{k\ell}||^2_{\textrm{max}}, \label{eq:interf}
\end{align}
where the upper bound in \eqref{eq:interf} is attained when both ${\bf Q}_k, {\bf Q}_\ell$ are fully-aligned rank-1 matrices. Since, by definition, $||{\bf V}_{k\ell}||_{\textrm{max}}\leq 1$, we can infer that at most ${\rm tr}\Big({\bf Q}_k {\bf Q}_\ell\Big)=M^2$. On the other extreme, we now consider the ideal case of no inter-user interference, that is ${\rm tr}\Big({\bf Q}_k {\bf Q}_\ell\Big)=0$, which requires the covariance matrices to have orthogonal support, i.e. ${\bf Q}_k {\bf Q}_\ell={\bf 0}$. This condition was utilized in a stream of papers \cite{Support1,Support2} to eliminate the effects of pilot contamination, by allocating pilots to users whose covariance matrices have nearly orthogonal support. However, as was articulated in \cite{Emil_PC}, the orthogonality support condition is very unlikely in practice. 
We now consider two intuitive scenarios for ${\bf Q}_1$, ${\bf Q}_2$ to examine the effect of the matrix rank on the amount of 
inter-user covariance  interference. 
\begin{align*}
&\textrm{Scenario 1:}~ {\bf Q}_1 = \left[ {\begin{array}{cc}
{\bf 1}_{M-D,M-D} & {\bf 0}_{M-D,D}\\
{\bf 0}_{D,M-D} & {\bf I}_{D}\\
\end{array}}  \right],~~{\bf Q}_2 = {\bf 1}_{M}\\
&\textrm{Scenario 2:}~ {\bf Q}_1 = \left[ {\begin{array}{cc}
{\bf 1}_{D,M-D} & {\bf 0}_{D,D}\\
{\bf 0}_{M-D,D} & {\bf I}_{M-D}\\
\end{array}}  \right],~~{\bf Q}_2 = {\bf 1}_{M},
\end{align*}
where ${\bf 1}_{m,n}$ is a $m\times n$ matrix full of 1's, and $D$ is a rank control parameter. In Fig.\ref{fig:traces}, we validate that
for Scenario 1 and low values of $D$, ${\bf Q}_1$ is rank-deficient and most, importantly, aligned with ${\bf Q}_2$. This is a catastrophic scenario that makes the inter-user covariance interference term scale with $M^2$. On the other extreme, for Scenario 2 and $D=1$, we have that 
${\bf Q}_1$ is full-rank and this makes  ${\rm tr}({\bf Q}_1{\bf Q}_2)/M^2$ approach zero. As $D$ increases, ${\bf Q}_1$  becomes more and more rank-deficient, yet, it never becomes fully aligned with ${\bf Q}_2$.

\begin{figure}
\centering\includegraphics[width=8.85cm]{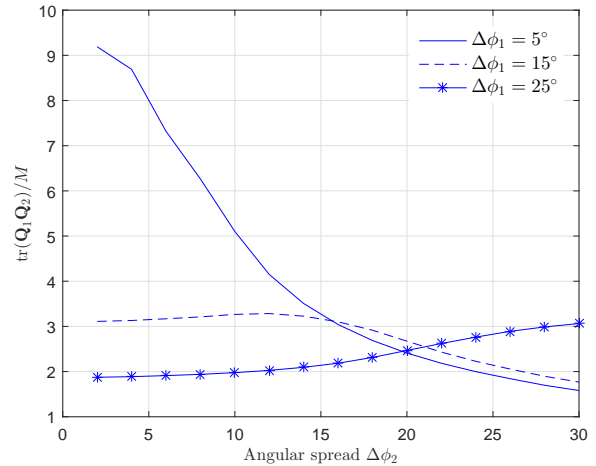}
\caption{Inter-user covariance interference term, ${\rm tr}({\bf Q}_1{\bf Q}_2)/M$ against the angular spread $\Delta\phi_2$ ($M=100, 
\phi_0^1=\pi/4, \phi_0^2=3\pi/4, d=1/2$).}
\label{fig:traces2}
\end{figure}

We will now investigate the case when ${\rm tr}\Big({\bf Q}_k {\bf Q}_\ell\Big)$ scales as ${\cal O}(M)$, which lies in the intersection of the previously investigated scenarios. To this end, we leverage Chebyshev's sum inequality and obtain the following lower bound:
\begin{align}
	{\rm tr}\Big({\bf Q}_k {\bf Q}_\ell\Big)&\geq\sum_{m=1}^M\sum_{n=1}^M|{\bf V}_{k\ell}(m,n)|^2\notag\\
	&=|| {\bf V}_{k\ell} ||_F^2=M.
	\label{eq:Lower}
\end{align}
Note that the lower bound in \eqref{eq:Lower} becomes exact if $\omega_{\ell,m}, \omega_{k,n}$ are all equal, such that
$\omega_{\ell,m}=1,\forall m=1,\ldots,M$ and $\omega_{k,n}=1, \forall n=1,\ldots,M$. To visualize this case, we now consider the one-ring model, 
for which, the $(i,j)$-th entry of ${\bf Q}_k$ is given by
\begin{align}
 [{\bf Q}_k]_{i,j}= \frac{1}{2\Delta\phi_{k}}\int_{-\Delta\phi_{k} + \phi_0^k}^{\Delta\phi_{k}+ \phi_0^k} e^{-j2\pi d(j-i) \sin(\phi_k)}d\phi_k,
\end{align}
where $\Delta\phi_{k}$ is the azimuth angular spread corresponding to the $k$-th user, $\phi_0^k$ is the nominal direction-of-arrival, while $d$ is the normalized antenna spacing as in \eqref{eq:LoS_resp}. Figure \ref{fig:traces2} compares the term, ${\rm tr}({\bf Q}_1{\bf Q}_2)/M$, 
against the angular spread $\Delta\phi_2$ for different values of $\Delta\phi_1$. The graph shows a number of interesting observations: First, 
when both ${\bf Q}_1,{\bf Q}_2$ are very rank-deficient (i.e. they have small $\Delta\phi_{k}$), the interference term is boosted. 
Surprisingly, when the angular spreads are high, e.g., $\Delta\phi_{k}>20^{\circ}$, this is not the ideal scenario since we notice a steady increase of ${\rm tr}({\bf Q}_1{\bf Q}_2)/M$. This implies that, from the perspective of minimum inter-user covariance interference, the best scenario is when one covariance matrix is full-rank and the other one is very rank-deficient.

\section{Conclusion}
The main motivation behind this work has been the inherent deficiencies of the Kronecker-type models.
For this reason, we invoked a generalized spatial correlation model, namely the Weichselberger model, which 
is able to effectively capture the average coupling between the spatial subchannels. Our analysis began by recasting the standard point-to-point Weichselberger model to a massive MIMO setup. We then derived closed-form expressions for the average desired signal and interference powers that are indispensable for a comprehensive performance analysis. In particular, we examined the two critical performance measures of any massive MIMO system, these are channel hardening and FP. Our analysis showcased the impact of a number of system parameters on these measures, such as the scaling behavior of the coupling vectors, rank of the covariance matrix, and angular spread. Our conclusions articulate that 
if the coupling vector has at least one entry that scales as ${\cal O}(M)$, channel hardening breaks down. Moreover, FP between two users is more pronounced whenever one user's covariance matrix is full-rank and the other one is very rank-deficient. As a final remark, in our future work, we will elaborate on the achievable spectral efficiency and how this is affected by model parameters.

\appendices
\section{Proof of Proposition \ref{Moments}}\label{app1}
We begin the proof by expanding the left-hand side of \eqref{eq:Signal_power1}, as follows:
\begin{align}
	&{\mathbb E}\Big[||{\bf h}_k||^4\Big]  = {\mathbb E}\Big[{\bf h}_k^H{\bf h}_k{\bf h}_k^H{\bf h}_k\Big] \notag\\
	&={\mathbb E}\Bigg[\left(\eta_k^2\bar{\bf {h}}_k^H\bar{\bf {h}}_k  +\eta_k\gamma_k\bar{\bf {h}}_k^H  \widetilde{\bf {h}}_k+ \eta_k\gamma_k\widetilde{\bf {h}}_k^H\bar{\bf {h}}_k +\gamma_k^2\widetilde{\bf {h}}_k^H\widetilde{\bf {h}}_k\right)^2 \Bigg]\notag\\
	&={\mathbb E}\Bigg[\left(\eta_k^2M + \eta_k\gamma_k\bar{\bf {h}}_k^H  \widetilde{\bf {h}}_k+ \eta_k\gamma_k\widetilde{\bf {h}}_k^H\bar{\bf {h}}_k+\gamma_k^2||\widetilde{\bf {h}}_k||^2\right)^2 \Bigg] \notag \\
	&=\eta_k^4M^2+\eta_k^2\gamma_k^2M{\mathbb E}\Big[||\widetilde{\bf {h}}_k||^2\Big]\notag\\
	&+2\eta_k^2\gamma_k^2{\mathbb E}\Big[\bar{\bf {h}}_k^H\widetilde{\bf {h}}_k\widetilde{\bf {h}}_k^H\bar{\bf {h}}_k^H\Big] 
	+\gamma_k^4{\mathbb E}\Big[||\widetilde{\bf {h}}_k||^4\Big],
	\end{align}
where we have removed all zero-mean terms from the cross-products. By noticing that ${\mathbb E}\Big[||\widetilde{\bf {h}}_k||^2\Big]={\rm tr}(\Lambdav_k)=M$ and that ${\mathbb E}\Big[||\widetilde{\bf {h}}_k||^4\Big]=\Big({\rm tr}(\Lambdav_k^2)+({\rm tr}\left(\Lambdav_k\right))^2\Big)$ \cite{Bjornson_Matthaiou,Tataria2017}, the proof follows after appropriate simplifications. The proof for \eqref{eq:Signal_power2} is trivial and therefore is omitted. 

\section{Proof of Proposition \ref{Moments2}}\label{app2}
The proof follows a similar line of reasoning as above. In particular, we have
\begin{align}
&{\mathbb E}\Big[\big|{\bf h}_k^H{\bf h}_\ell\big|^2\Big] = {\mathbb E}\Big[{\bf h}_k^H{\bf h}_\ell{\bf h}_\ell^H{\bf h}_k\Big]\notag\\
&={\mathbb E}\Big[\big(\eta_k \bar{\bf {h}}_k^H+\gamma_k \widetilde{\bf {h}}_k^H\big)\big(\eta_\ell \bar{\bf {h}}_\ell+\gamma_\ell \widetilde{\bf {h}}_\ell\big)\\
&\hspace{15pt}\times\big(\eta_\ell \bar{\bf {h}}_\ell^H+\gamma_\ell \widetilde{\bf {h}}_\ell^H\big)\big(\eta_k \bar{\bf {h}}_k+\gamma_k \widetilde{\bf {h}}_k\big)\Big]\notag\\
&= \eta_k^2\eta_\ell^2\big|\bar{\bf {h}}_k^H\bar{\bf {h}}_\ell\big|^2
+\gamma_k^2\gamma_\ell^2{\mathbb E}\Big[\widetilde{\bf {h}}_k^H\widetilde{\bf {h}}_\ell\widetilde{\bf {h}}_\ell^H \widetilde{\bf {h}}_k \Big]\notag\\
&+\eta_k^2\gamma_\ell^2{\mathbb E}\Big[\bar{\bf {h}}_k^H \widetilde{\bf {h}}_\ell\widetilde{\bf {h}}_\ell^H \bar{\bf {h}}_k \Big]+\eta_\ell^2\gamma_k^2{\mathbb E}\Big[\bar{\bf {h}}_\ell^H \widetilde{\bf {h}}_k\widetilde{\bf {h}}_k^H \bar{\bf {h}}_\ell \Big]
\end{align}
and the proof concludes by recalling the definition of ${\bf Q}_k$ and ${\bf Q}_\ell$ from \eqref{eq:Covariance} and some basic algebra. 
\label{app:LARGE}

\end{document}